\documentclass[11pt]{article}

\usepackage[margin=1in]{geometry}
\usepackage{graphicx}
\usepackage{booktabs}
\usepackage{amsmath,amssymb}
\usepackage{siunitx}
\usepackage{array}
\usepackage{multirow}
\usepackage{float}
\usepackage{setspace}
\usepackage[hidelinks]{hyperref}
\usepackage{caption}
\usepackage{subcaption}
\usepackage{microtype}
\usepackage{tabularx}
\usepackage[backend=biber,style=numeric,sorting=none]{biblatex}
\title{Twist-Tuned Bilayer Metasurface for 3~T MRI}

\author{
Ingrid Torres$^{1}$, Raquel Rodriguez$^{2}$, Robert W. Laird$^{2,3}$, Angela R. Laird$^{2,3}$,\\ and Alex Krasnok$^{1,*}$\\[4pt]
\small $^{1}$Department of Electrical and Computer Engineering, Florida International University,\\
\small Miami, FL 33174, USA\\
\small $^{2}$Center for Imaging Science, Florida International University, Miami, FL 33199, USA\\
\small $^{3}$Department of Physics, Florida International University, Miami, FL 33199, USA\\
\small $^{*}$Corresponding author: akrasnok@fiu.edu
}
\date{}

\begin{document}
\maketitle
\doublespacing

\begin{abstract}
Magnetic resonance imaging (MRI) can see deep inside the body without ionizing radiation, but image quality depends strongly on how well the radio-frequency field is controlled. Passive resonant pads and metasurfaces can help, yet they often lose their tuning when they are placed next to water-rich tissue or tissue-like materials. Here we show a simple way to bring such a device back into tune. We built a bilayer metasurface made of two aluminum wire arrays. One layer can rotate relative to the other, and the gap between the two layers can also be adjusted. Bench measurements show that adding a controlled water load shifts the resonance to lower frequency by about \SIrange{4.2}{11.4}{\mega\hertz}. Rotating the layers shifts it back by about \SIrange{13.2}{14.9}{\mega\hertz}, which is much stronger than changing the gap alone. One loaded setting lands essentially at the proton frequency used in \SI{3}{\tesla} MRI. In a proof-of-concept scan on a clinical \SI{3}{\tesla} system, the metasurface made internal features in a structured pineapple phantom easier to see than in a substrate-only control. These results show that a passive MRI metasurface can be tuned after fabrication and retuned under load using geometry alone, opening a practical route to simple adjustable RF accessories for MRI.
\end{abstract}

\section{Introduction}

Magnetic resonance imaging (MRI) is one of the most important tools in modern medicine. It can reveal soft tissue deep inside the body without ionizing radiation, and it is used across neurology, oncology, cardiology, and musculoskeletal imaging. MRI is powerful not only because it produces anatomical images, but also because it can probe structure, composition, and function in the same platform. That broad reach is why improvements in MRI hardware matter: even modest gains in local field control can translate into clearer images, stronger signal, shorter scans, or more reliable imaging in anatomically difficult regions.

At the same time, MRI is fundamentally an electromagnetic problem. In proton MRI, the static magnetic field $B_0$ sets the Larmor frequency $f_{\mathrm L}$, which is the frequency at which the nuclear spins respond most strongly. The transmit field $B_1^+$ excites those spins, and the receive field $B_1^-$ determines how efficiently the returning signal is detected \cite{HoultRichards1976,Glover1985,Roemer1990,Hoult2000,HoultReciprocity2011,Li2024NonlinearPMRI}. For $^1$H imaging at \SI{3}{\tesla}, the proton Larmor frequency is
\begin{equation}
f_{\mathrm L}=\frac{\gamma}{2\pi}B_0=\SI{127.73}{\mega\hertz},
\end{equation}
where $\gamma$ is the proton gyromagnetic ratio \cite{NISTGamma}. At this frequency, conductors with decimeter-to-meter length scales can interact strongly with the RF field. Their response depends not only on their own geometry, but also on the nearby sample, the RF coil, and any water-rich, lossy object placed next to them. As a result, a structure that is tuned correctly in air can shift away from the desired frequency once it is brought near tissue or a tissue-like load.

\begin{figure}[t]
\centering
\includegraphics[width=0.8\linewidth]{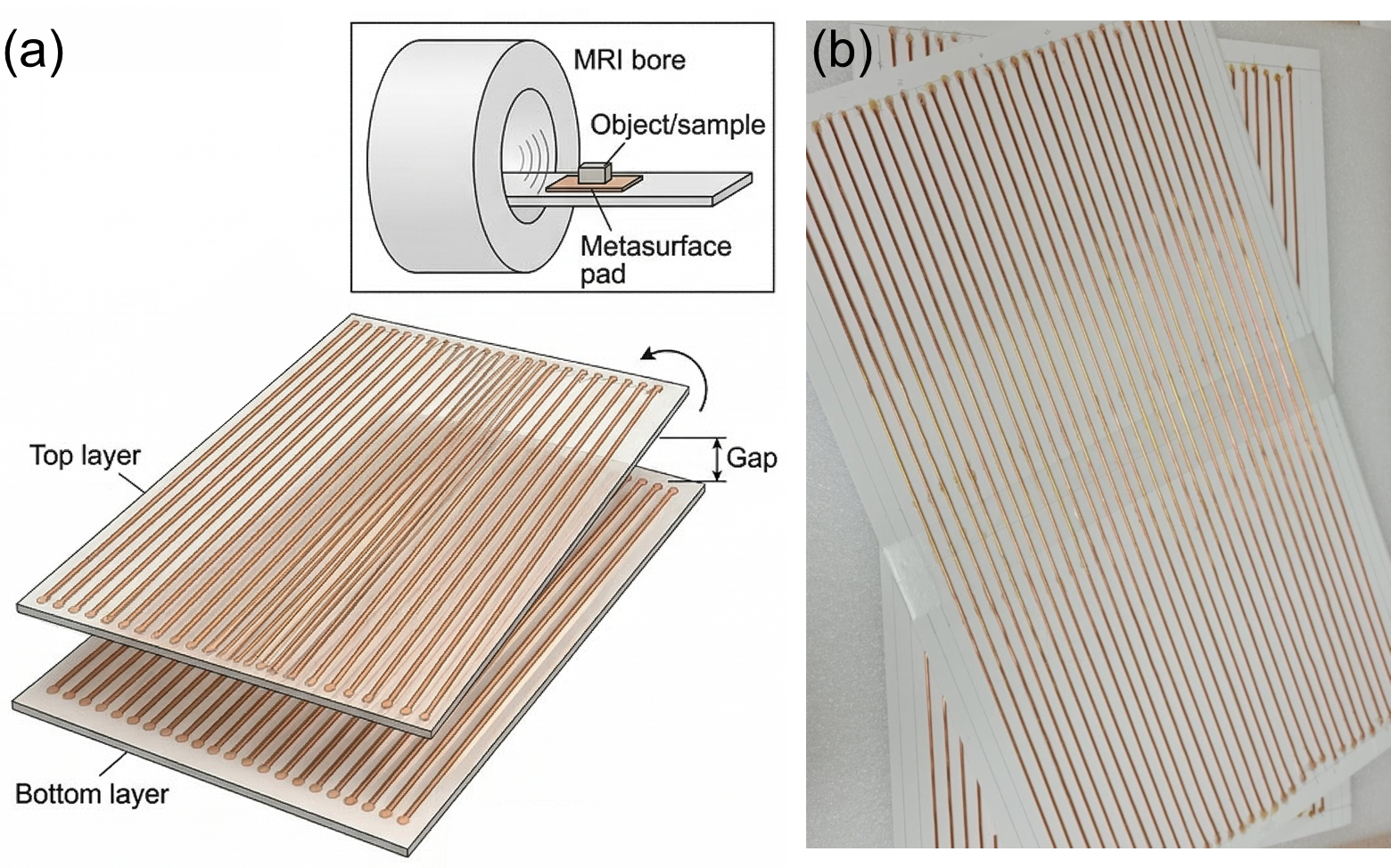}
\caption{Bilayer wire metasurface used in this study. (a) Conceptual drawing of the device. The two wire layers are separated by an adjustable gap $G$ and rotated by a relative in-plane angle $\phi$. The inset shows the intended placement of the pad beneath the sample inside the MRI bore. (b) Photograph of the prototype used for the bench and MRI measurements.}
\label{fig:conceptphoto}
\end{figure}

Passive RF shimming is an appealing way to address that challenge. High-permittivity, or high-dielectric-constant (HDC), pads are the most established example and can improve transmit efficiency, receive sensitivity, and image quality in both research and clinical settings \cite{Sica2020,Webb2022NovelMaterials}. Resonant metasurfaces offer a different but related strategy. A metasurface is an engineered sheet made of subwavelength elements that reshape the surrounding electromagnetic field. In MRI, such structures can locally enhance the RF magnetic field and redirect it toward a region of interest \cite{Slobozhanyuk2016,Webb2022NovelMaterials}. Recent work has shown a wide range of approaches, including wire metasurfaces, ceramic resonators, artificial dielectrics, nonlinear self-detuning designs, mechanically deformable structures, wearable systems, and electronically tunable platforms \cite{Shchelokova2018,Shchelokova2020,Vorobyev2020,Stoja2021,Wu2022Auxetics,Zhu2024Wearable,Koloskov2024,Koloskov2025Fetal,Chen2025Tunable}. Together, these studies make one point clear: passive and semi-passive engineered structures can improve MRI. The harder problem is making them practical.

The practical difficulty is detuning under load. A resonant insert may work well in isolation, yet shift by several megahertz when it is placed next to a water-rich object. That shift can reduce or even eliminate the intended benefit. Existing strategies only partly solve this problem. HDC pads are effective but depend strongly on placement and material choice \cite{Sica2020}. Nonlinear or electronically tunable metasurfaces can adapt to changing conditions, but they introduce extra components, biasing, or circuit complexity \cite{Stoja2021,Chen2025Tunable,Li2024NonlinearPMRI}. Flexible and wearable structures conform better to anatomy, but their resonance is still often set mainly by the original design \cite{Zhu2024Wearable,Koloskov2024,Koloskov2025Fetal}. What is still missing is a simple passive insert that can be tuned after fabrication and then retuned again after loading, without active electronics.

In this work, we address that gap with a bilayer wire metasurface made of two closely spaced aluminum wire arrays, shown schematically in Fig.~\ref{fig:conceptphoto}(a) and photographed in Fig.~\ref{fig:conceptphoto}(b). The device has two mechanical control variables: the relative in-plane rotation angle $\phi$ and the interlayer gap $G$. The physical idea is straightforward. The two layers behave as coupled resonators. Rotating one layer relative to the other changes their overlap and symmetry, which can shift the hybrid resonance strongly. Changing the gap also changes the coupling, but more gently. In that sense, rotation acts as the coarse tuning knob and gap acts as the fine tuning knob. This picture is consistent with our earlier lower-frequency study of the same bilayer architecture, where twist-driven hybridization produced strong tuning at small gap and much weaker tuning at larger separation \cite{Torres2026TwistTuned}.

We then bring this bilayer concept to the clinical proton band at \SI{3}{\tesla} and test it in two steps. First, we perform calibrated bench measurements to determine how dielectric loading shifts the resonance and how strongly each mechanical control can move it back. These measurements show that a controlled water load consistently shifts the resonance downward, that rotation provides much stronger tuning authority than gap, and that the loaded state can be brought essentially onto the \SI{3}{\tesla} proton frequency. Second, we test the device in a proof-of-concept scanner experiment using a structured, water-rich phantom and compare it with a substrate-only control. The metasurface produces clearer internal structure, linking the bench tuning result to a visible imaging consequence.

The importance of these results is broader than one prototype. They show that passive MRI metasurfaces do not need to be fixed once fabricated. They can be adjusted after fabrication and retuned for different loading conditions using geometry alone. That capability opens a path toward simpler and more practical RF accessories for MRI: reusable research platforms, low-complexity patient-adaptive inserts, and future anatomy-specific or flexible metasurfaces that can be matched to the imaging task without active electronics. In short, the work shifts the focus from making a resonant insert to keeping it useful in the setting where MRI actually operates.

\section{Results and Discussion}

\subsection{Bench measurements identify the loading-induced shift}

\begin{figure}[t]
\centering
\includegraphics[width=0.98\linewidth]{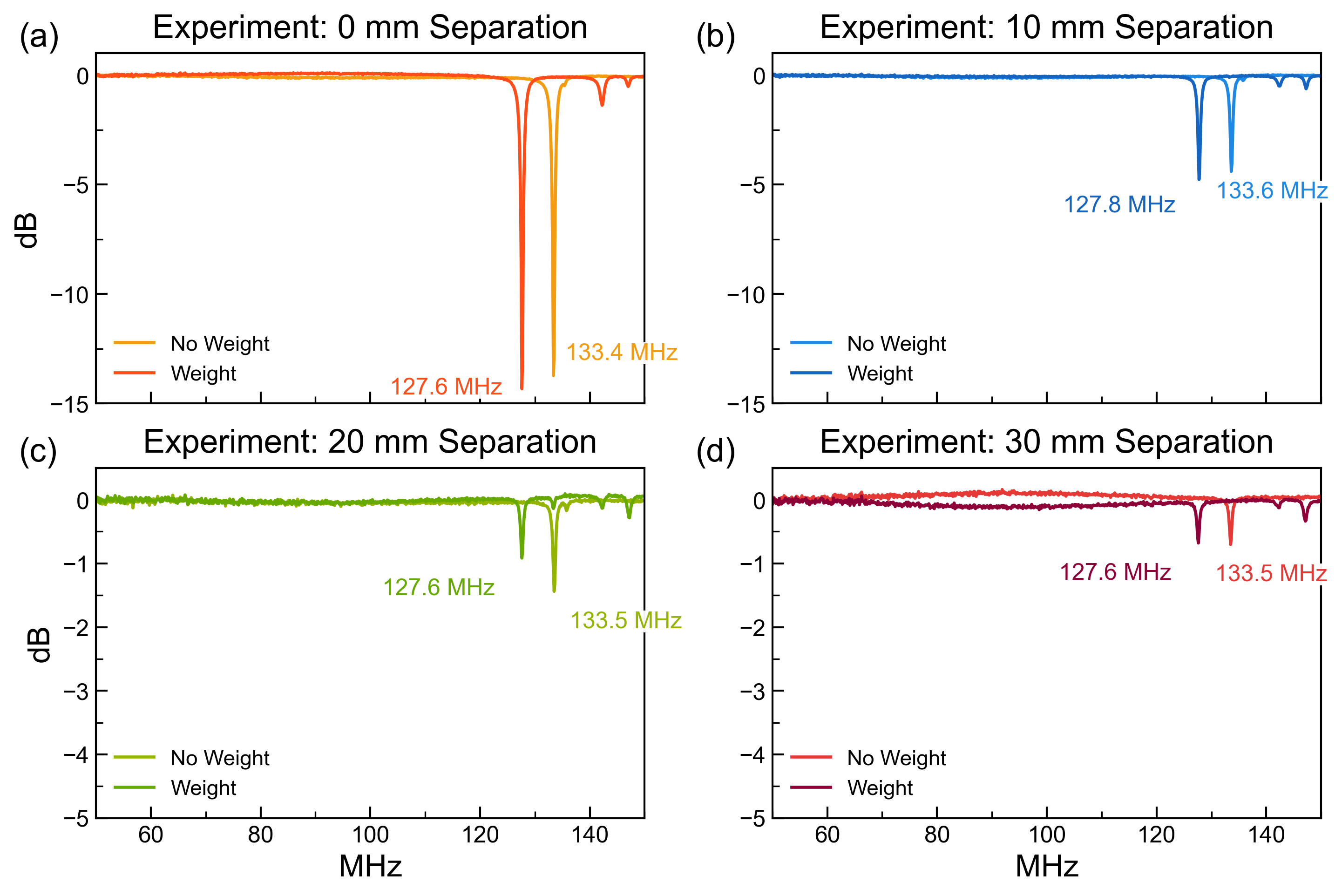}
\caption{Representative calibrated reflection spectra at different interlayer gaps. Normalized versions of calibrated $S_{11}$ traces are shown for Family~C, the gap-sweep family that remains nearest the proton band after loading, at gaps of (a)~\SI{0}{\milli\meter}, (b)~\SI{10}{\milli\meter}, (c)~\SI{20}{\milli\meter}, and (d)~\SI{30}{\milli\meter}. In every panel, the water-loaded state shifts the tracked resonance to lower frequency than the unloaded state.}
\label{fig:gapvna}
\end{figure}

Bench characterization used one-port loop-coupled measurements of the complex reflection coefficient $S_{11}$, where $S_{11}$ quantifies the fraction of the applied RF signal reflected back to the vector network analyzer (VNA). Five archived gap sweeps, labeled Families~A--E, were recorded at nominal gaps of \SIlist{0;10;20;30}{\milli\meter}. A separate minimal-gap scan was acquired at two reproducible indexed rotation states, denoted nominal \SI{10}{\degree} and nominal \SI{20}{\degree}. These families correspond to repeatable nearby mechanical states rather than optically calibrated absolute angles, as discussed in Supplementary Section~S4. A controlled dielectric load was introduced by placing two water-filled beakers on the upper layer. Throughout the paper, \emph{unloaded} denotes the state without the beakers and \emph{loaded} denotes the otherwise identical state with the beakers in place. The spectra in Fig.~\ref{fig:gapvna}(a)--(d) and Fig.~\ref{fig:rotationvna}(a),(b) are normalized for visual comparison, but the frequencies reported below were extracted from the calibrated traces by consistent branch tracking across matched states.

Across all five families, dielectric loading red-shifted the tracked resonance. Figure~\ref{fig:gapvna}(a)--(d) shows representative spectra from Family~C, the gap-sweep family that remains nearest the proton band after loading. In every panel, the loaded resonance lies below the unloaded one. The full archive in Supplementary Table~S4 gives load-induced shifts from \SI{4.17}{\mega\hertz} to \SI{11.40}{\mega\hertz}. The loaded state nearest the proton band occurs in Family~C at \SI{10}{\milli\meter} gap, where the tracked resonance is \SI{127.77}{\mega\hertz}, only \SI{0.04}{\mega\hertz} above $f_{\mathrm L}$ at \SI{3}{\tesla}. Mechanical retuning is therefore sufficient to recover the operating band under dielectric loading.

A compact estimate of the net reactive change follows from the resonance relation
\begin{equation}
f=\frac{1}{2\pi\sqrt{LC}},
\end{equation}
where $L$ and $C$ are the effective inductance and capacitance of the tracked mode. For matched loaded and unloaded states,
\begin{equation}
\frac{(LC)_{\mathrm{loaded}}}{(LC)_{\mathrm{unloaded}}}\approx \left(\frac{f_{\mathrm{unloaded}}}{f_{\mathrm{loaded}}}\right)^2.
\label{eq:LCratio}
\end{equation}
Using the extracted frequencies in Supplementary Table~S4, the inferred ratio $(LC)_{\mathrm{loaded}}/(LC)_{\mathrm{unloaded}}$ spans 1.069--1.192, corresponding to an increase of \SIrange{6.9}{19.2}{\percent} in the effective reactive product, with a median ratio of 1.092. This estimate does not separate inductive and capacitive contributions, but it sets the size of the correction that a practical tuning mechanism must provide.

\subsection{Gap trims the resonance, whereas rotation recovers the band}

The gap scan shows that $G$ acts as a trim control rather than the primary tuning variable. In Family~A, the unloaded resonance moves from \SI{129.44}{\mega\hertz} at \SI{0}{\milli\meter} to \SI{127.55}{\mega\hertz} at \SI{30}{\milli\meter}, which corresponds to an average slope of about \SI{-0.063}{\mega\hertz\per\milli\meter}. In Family~D, the same quantity changes by only \SI{0.31}{\mega\hertz} across the full \SI{30}{\milli\meter} range, or about \SI{-0.010}{\mega\hertz\per\milli\meter}. The near-target loaded family is even more revealing: in Family~C, the loaded resonance remains within \SI{0.17}{\mega\hertz} across all four gap states and reaches its closest approach to the proton band at \SI{10}{\milli\meter}, where $f_{\mathrm{loaded}}=\SI{127.77}{\mega\hertz}$. Gap is therefore useful for final placement of the resonance, but it does not rescue a strongly detuned state.

\begin{figure}[t]
\centering
\includegraphics[width=0.98\linewidth]{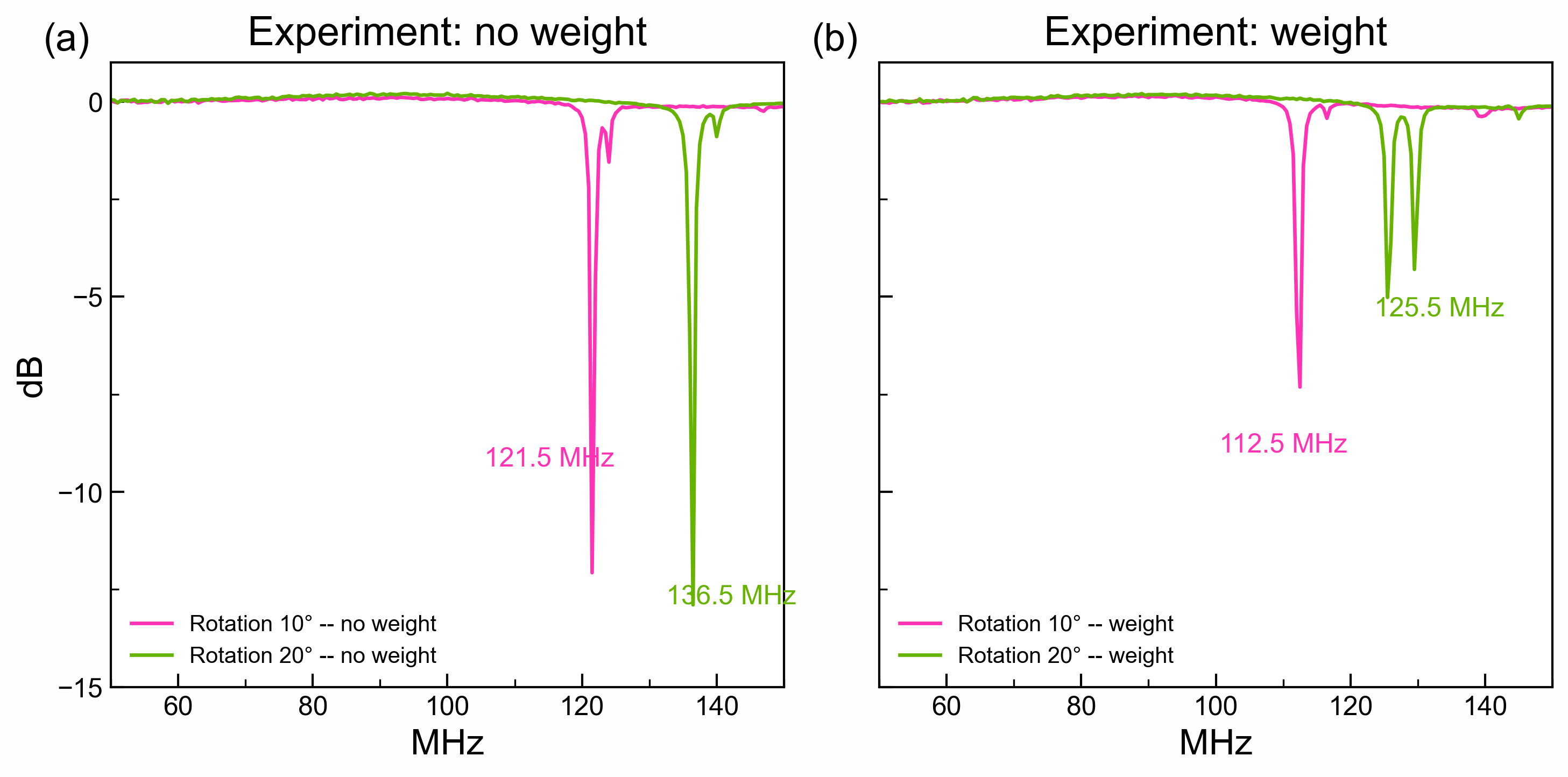}
\caption{Rotation-driven coarse tuning at the minimal-gap setting. Normalized versions of calibrated $S_{11}$ traces are shown for the two indexed rotation states in (a) the unloaded case and (b) the water-loaded case. The tracked resonance shifts by \SI{14.92}{\mega\hertz} without the added water load and by \SI{13.24}{\mega\hertz} with the added water load, showing that rotation is the dominant tuning variable.}
\label{fig:rotationvna}
\end{figure}

At minimal gap, the role of rotation is much stronger. Figure~\ref{fig:rotationvna}(a) compares the unloaded spectra for the two indexed states, and Fig.~\ref{fig:rotationvna}(b) shows the corresponding loaded spectra. Without the water load, the tracked resonance moves from \SI{121.53}{\mega\hertz} at nominal \SI{10}{\degree} to \SI{136.45}{\mega\hertz} at nominal \SI{20}{\degree}. With the load present, it moves from \SI{112.37}{\mega\hertz} to \SI{125.61}{\mega\hertz}. The rotation-driven shift is therefore \SI{14.92}{\mega\hertz} in the unloaded state and \SI{13.24}{\mega\hertz} in the loaded state. Expressed as local between-state sensitivities, these values correspond to about \SI{1.49}{\mega\hertz\per\degree} and \SI{1.32}{\mega\hertz\per\degree}, respectively. Because the two indexed states were set mechanically rather than by optical angle metrology, those sensitivities should be read as local estimates, not universal device constants. The conclusion is simpler: changing rotation moves the resonance by an order of magnitude more than changing gap over the tested range. Rotation provides the coarse correction needed to recover the band; gap trims the final operating point.

\begin{table}[H]
\centering
\caption{Main quantitative tuning results from the calibrated bench measurements. The full resonance-extraction log is given in Supplementary Table~S4, and the family mapping and measurement context are summarized in Supplementary Sections~S1 and S4.}
\label{tab:tuning_summary}
\small
\begin{tabular}{@{}p{6.6cm}p{7.1cm}@{}}
\toprule
Quantity & Result \\
\midrule
Load-induced resonance shift across all measured gap sweeps & \SIrange{4.17}{11.40}{\mega\hertz} downward shift \\
Median effective reactive-product ratio from Eq.~\eqref{eq:LCratio} & $(LC)_{\mathrm{loaded}}/(LC)_{\mathrm{unloaded}} \approx 1.092$ \\
Loaded state closest to the \SI{3}{\tesla} proton frequency & \SI{127.77}{\mega\hertz} at \SI{10}{\milli\meter} gap in Family~C, within \SI{0.04}{\mega\hertz} of $f_{\mathrm L}$ \\
Largest observed gap-driven change over \SIrange{0}{30}{\milli\meter} & \SI{1.89}{\mega\hertz} in unloaded Family~A \\
Gap span of the loaded near-target family (Family~C) over \SIrange{0}{30}{\milli\meter} & \SI{0.17}{\mega\hertz} \\
Rotation-driven shift at minimal gap & \SI{14.92}{\mega\hertz} unloaded; \SI{13.24}{\mega\hertz} loaded \\
Approximate local rotation sensitivity between indexed nominal \SI{10}{\degree} and nominal \SI{20}{\degree} states & \SI{1.49}{\mega\hertz\per\degree} unloaded; \SI{1.32}{\mega\hertz\per\degree} loaded \\
\bottomrule
\end{tabular}
\end{table}

\subsection{A coupled-resonator picture explains the tuning hierarchy}

A minimal phenomenological model captures both the loading red shift and the different roles of $\phi$ and $G$. Each wire layer is represented by an effective self-inductance $L_0$ and self-capacitance $C_0$. The interlayer interaction is described by a mutual inductance $M(\phi,G)$ and a mutual capacitance $C_m(\phi,G)$, while a nearby dielectric object contributes an additional capacitance $\Delta C_{\mathrm{obj}}$ through the fringing electric field. The two hybridized angular frequencies, $\omega_{\pm}=2\pi f_{\pm}$, can then be written approximately as
\begin{equation}
\omega_{\pm}(\phi,G)\approx \frac{1}{\sqrt{\left[L_0\pm M(\phi,G)\right]\left[C_0\pm C_m(\phi,G)+\Delta C_{\mathrm{obj}}\right]}}.
\label{eq:hybrid_branches}
\end{equation}
Supplementary Section~S6 expands this phenomenological picture and its limits.

Equation~\eqref{eq:hybrid_branches} immediately explains the sign of the loading shift. A water-rich object increases the electric energy stored in the near field and therefore increases the effective capacitance sampled by the mode. The resonance frequency must then decrease, exactly as observed in Fig.~\ref{fig:gapvna}(a)--(d) and Fig.~\ref{fig:rotationvna}(b). The same equation also explains the tuning hierarchy. Changing $G$ mainly scales a given coupling configuration and therefore nudges a selected mode, whereas changing $\phi$ alters overlap symmetry and can move the hybrid branch itself by many megahertz. This is the same hierarchy reported in the lower-frequency bilayer study, where strong continuous tuning appeared only in the small-gap regime and collapsed rapidly as the layers were separated \cite{Torres2026TwistTuned}. The \SI{3}{\tesla} measurements reported here follow the same logic in the clinical proton band.

\subsection{A proof-of-concept \SI{3}{\tesla} scan shows an imaging consequence}

\begin{figure}[t]
\centering
\includegraphics[width=0.98\linewidth]{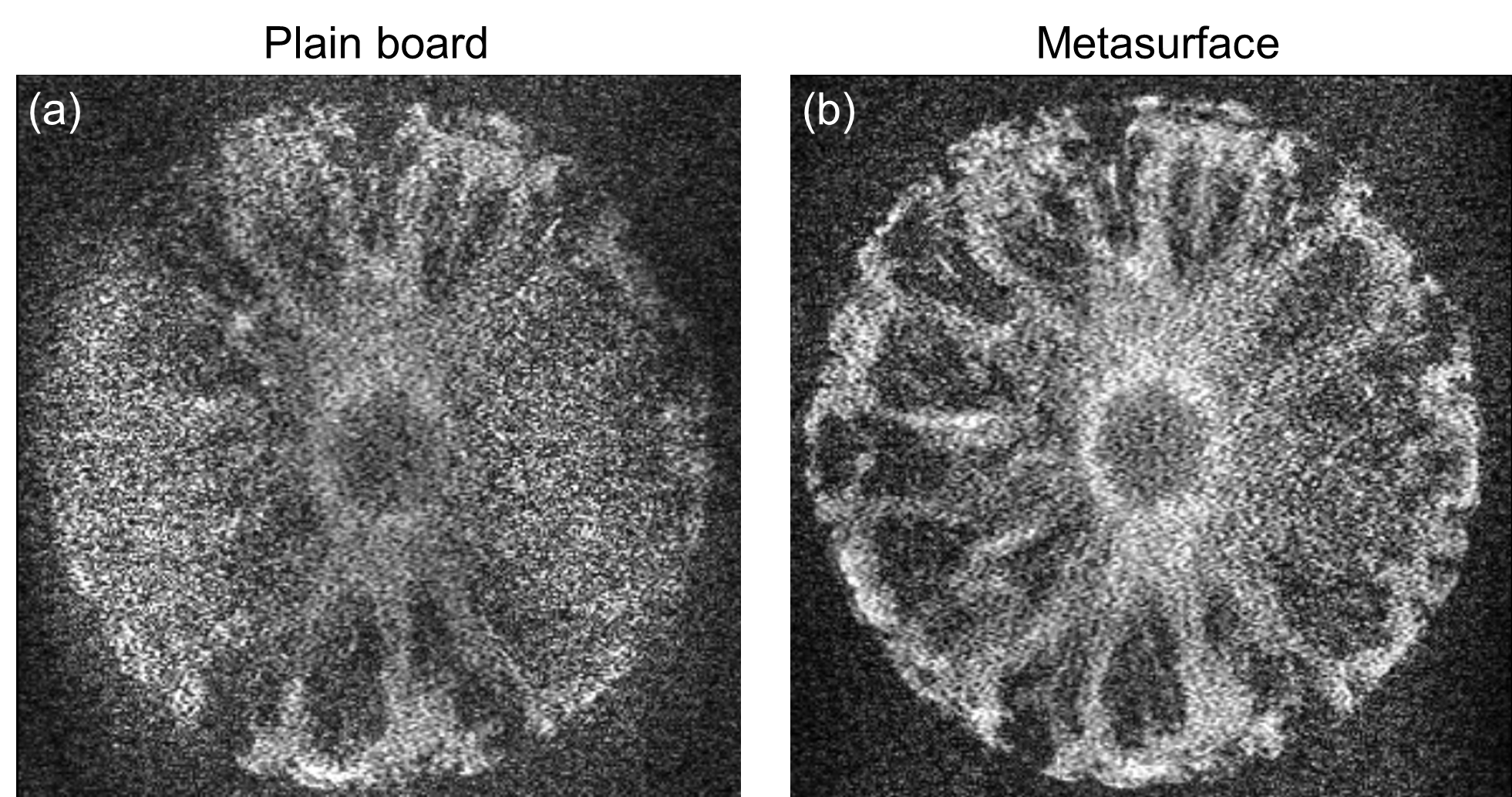}
\caption{MRI proof-of-concept demonstration on a structured, water-rich fruit phantom. Representative images from the matched scan pair acquired with the posterior/spine receive array disabled are shown for (a) the substrate-only control and (b) the metasurface-supported scan at the nominal \SI{10}{\degree} indexed state. The metasurface condition shows clearer internal radial structure and a sharper boundary between the bright outer region and the darker core.}
\label{fig:mri}
\end{figure}

MRI measurements were performed on a clinical \SI{3}{\tesla} system using a \SI{1}{\centi\meter}-thick pineapple slice as a structured, water-rich fruit phantom. The pineapple was chosen because it offers strong internal radial features and a clear boundary between the brighter outer region and the darker central core. The metasurface was placed beneath the sample, as indicated in the inset of Fig.~\ref{fig:conceptphoto}(a). The matched comparison in Fig.~\ref{fig:mri}(a),(b) used the metasurface at the nominal \SI{10}{\degree} indexed state and a substrate-only control, with the posterior/spine receive array disabled and all scanner settings held fixed between the two scans. The preserved image set contains eight slices of thickness \SI{2.0}{\milli\meter}, matrix size $288\times 288$, and in-plane pixel spacing \SI{0.444}{\milli\meter}. The scan summary is given in Supplementary Table~S3, and additional matched slices are shown in Supplementary Fig.~S2.

The difference between Fig.~\ref{fig:mri}(a) and Fig.~\ref{fig:mri}(b) is visible without post-processing: with the metasurface present, the radial fibrous features are better separated and the ring-like boundary around the central core is sharper. To support that qualitative reading without over-claiming, the images were exported in DICOM (Digital Imaging and Communications in Medicine) format, rigidly registered, and analyzed on matched crops. Four descriptive metrics were used: Tenengrad sharpness, defined here as the mean squared Sobel-gradient magnitude; variance of the Laplacian, a second-derivative edge measure; high-frequency energy ratio, the fraction of Fourier-domain energy within a fixed high-frequency mask; and root-mean-square (RMS) contrast, the standard deviation of grayscale intensity about its mean. The explicit formulas and workflow are given in Supplementary Section~S5.

The results are listed in Table~\ref{tab:postproc_metrics}. All three edge-sensitive metrics favor the metasurface condition: Tenengrad sharpness increases by \SI{45.4}{\percent}, variance of the Laplacian by \SI{25.0}{\percent}, and high-frequency energy ratio by \SI{43.5}{\percent}. RMS contrast changes only weakly and decreases by \SI{3.7}{\percent}. That pattern is important. If the difference between Fig.~\ref{fig:mri}(a) and Fig.~\ref{fig:mri}(b) were only a global brightness rescaling, the four metrics would be expected to move more uniformly. Instead, the largest changes occur in quantities that respond specifically to edges and fine structure, which matches the visual impression of improved structural delineation.

\begin{table}[t]
\centering
\caption{Registered image-structure metrics for the matched MRI comparison in Fig.~\ref{fig:mri}(a),(b). The registration workflow and metric definitions are given in Supplementary Section~S5. These quantities support comparison of structural delineation, but they are not formal measurements of scanner spatial resolution.}
\label{tab:postproc_metrics}
\small
\begin{tabular}{@{}lccc@{}}
\toprule
Metric & Metasurface & Substrate-only control & Change (\%) \\
\midrule
Tenengrad sharpness & 1.023 & 0.704 & +45.4 \\
Variance of Laplacian & 0.323 & 0.259 & +25.0 \\
High-frequency energy ratio & 0.319 & 0.222 & +43.5 \\
RMS contrast & 0.240 & 0.249 & -3.7 \\
\bottomrule
\end{tabular}
\end{table}

The bench and scanner datasets answer different questions. The bench data establish the central device-physics result: the bilayer can be mechanically retuned back toward the proton band under a controlled dielectric load. The MRI data ask whether a tuned bilayer architecture changes image appearance under scanner conditions in the expected direction. Within the limits of the present dataset, the answer is yes. The scanner evidence is therefore appropriate as proof of concept, not as a calibrated performance benchmark. Absolute SNR, transmit efficiency, SAR, and scanner spatial resolution would require dedicated standardized measurements and are outside the scope of the present study \cite{ASTMF2182,NEMA_MS1,NEMA_MS6,NEMA_MS9}.

The design rule that emerges is simple. In this class of passive bilayer wire resonators, dielectric loading red-shifts the resonance, rotation provides the authority needed to recover the band, and gap trims the final placement of the resonance once the band has been recovered. That is the main result of the paper. Against the wider literature summarized in Supplementary Table~S2, the present work does not compete by using nonlinear components, variable capacitors, or complex three-dimensional architectures. Its advance is the combination of reversibility, geometric simplicity, and direct operation in the clinical \SI{3}{\tesla} proton band.

\section{Conclusions}

A passive resonant insert is useful at \SI{3}{\tesla} only if it can be brought back to the proton band after loading. The bilayer wire metasurface studied here does that mechanically. Across the preserved bench archive, water loading always red-shifted the tracked resonance, with shifts of \SIrange{4.17}{11.40}{\mega\hertz}. At minimal gap, rotation between the two indexed states provided the needed correction, moving the tracked branch by \SI{14.92}{\mega\hertz} in the unloaded case and by \SI{13.24}{\mega\hertz} in the loaded case, whereas gap changed the final frequency only modestly. The loaded state nearest the proton band occurred at \SI{127.77}{\mega\hertz}, showing that resonance recovery is possible without active electronics or lumped tuning elements. The scanner experiment gives that bench result an imaging consequence. In the matched \SI{3}{\tesla} comparison, the metasurface produced clearer internal structure in a structured, water-rich fruit phantom and higher edge-sensitive descriptive metrics than the substrate-only control. These data are properly interpreted as proof of concept rather than as standardized measurements of B$_1^+$, SNR, SAR, or resolution. For this bilayer class, the design rule is clear: dielectric loading red-shifts the resonance, rotation recovers the band, and gap trims the operating point. That combination of post-fabrication tunability, passive load compensation, and direct operation in the clinical proton band makes bilayer wire metasurfaces a practical platform for simple MRI RF accessories.

\section*{Acknowledgments}
AK acknowledges financial support from the U.S. Department of Energy (DoE) and the U.S. Air Force Office of Scientific Research (AFOSR).

\section*{Competing interests}
The authors declare no competing interests.

\section*{Data availability}
The datasets supporting this study are available from the corresponding author upon reasonable request.

\printbibliography

\end{document}